# Coherent control of single spins in silicon carbide at room temperature

Matthias Widmann[1], Sang-Yun Lee[1*], Torsten Rendler[1], Nguyen Tien Son[2], Helmut Fedder[1], Seoyoung Paik[1], Li-Ping Yang[3], Nan Zhao[3], Sen Yang[1], Ian Booker[2], Andrej Denisenko[1], Mohammad Jamali[1], Seyed Ali Momenzadeh[1], Ilja Gerhardt[1], Takeshi Ohshima[4], Adam Gali[5,6], Erik Janzén[2], Jörg Wrachtrup[1]

1. 3rd Institute of Physics and Research Center SCOPE, University of Stuttgart, Pfaffenwaldring 57, 70569 Stuttgart, Germany
2. Department of Physics, Chemistry and Biology, Linköping University, SE-58183 Linköping, Sweden
3. Beijing Computational Science Research Center, Beijing 100084, China
4. Japan Atomic Energy Agency, Takasaki, Gunma 370-1292, Japan
5. Wigner Research Centre for Physics, Hungarian Academy of Sciences, P.O. Box 49, H-1525, Budapest, Hungary
6. Department of Atomic Physics, Budapest University of Technology and Economics, Budafoki út 8, H-1111, Budapest, Hungary

*e-mail: s.lee@physik.uni-stuttgart.de

**Spins in solids are cornerstone elements of quantum spintronics[1]. Leading contenders such as defects in diamond[2–5], or individual phosphorous dopants in silicon[6] have shown spectacular progress but either miss established nanotechnology or an efficient spin-photon interface. Silicon carbide (SiC) combines the strength of both systems[5]: It has a large bandgap with deep defects[7–9] and benefits from mature fabrication techniques[10–12]. Here we report the characterization of photoluminescence and optical spin polarization from single silicon vacancies in SiC, and demonstrate that single spins can be addressed at room temperature. We show coherent control of a single defect spin and find long spin coherence time under ambient conditions. Our study provides evidence that SiC is a promising system for atomic-scale spintronics and quantum technology.**

The unique properties of SiC such as high thermal conductivity, high breakdown voltage, and wide bandgap has made it one of the most prominent materials for high power and high temperature electronic devices[13]. Various mature fabrication methods originating from silicon technology have been applied to SiC and fabrication of sophisticated nanostructures has been demonstrated[11]. By combining this with the unique properties of SiC, i.e. a wide bandgap for broadband optical access across the visible to infrared, SiC also has become a promising platform for nanophotonics[12]. Most recently, defects in SiC have attracted attention as potential host for quantum spin systems. Its weak spin-orbit coupling gives potential for long spin relaxation times[5,9,14]. Some defect centers sustain long relaxation times even at room temperature[9,14] similar to color centers in diamond[3,15]. Various defects show photoluminescence (PL) ranging from the visible to infrared which allows optical and even electrical detection of spin states[8,9,16] thanks to efficient p- and n-doping[10]. The existence of 250 known polytypes and diversity in defect spins supports the engineering of spin and optoelectronics properties of defect centers for various applications[17]. Neutral divacancy spins for example have revealed a high sensitivity to electric fields and also defects for temperature sensing have been found[18]. Its outstanding electromechanical properties[19] together with the capability of electrical detection[10] and electrical driving of spin resonance[20] may also pave new ways for scalable quantum devices. One important challenge towards spintronics and QIP application is the isolation of single spin defects. The single carbon antisite-vacancy pair ($C_{Si}V_C$) has been successfully created as bright single photon sources, however detection of its spin state has not been shown[21]. Here we use high purity SiC material to demonstrate single spin detection on single silicon vacancy defects in 4H-SiC ($T_{V2}$ centers[8]; see Fig. 1a).

A first key step is the controlled creation of $T_{V2}$ centers in a pure SiC crystal at the desired low density suitable for single defect detection. For this, we used a commercially available high purity SiC wafer and created $T_{V2}$ centers by 2 MeV electron irradiation without post-annealing. We characterized samples in a home-built confocal microscope (730 nm laser excitation). Figure 2a shows the PL emission spectrum of a single defect identified in a fluorescence image. The emission spectrum is similar to the characteristic fluorescence of a silicon vacancy defect ensemble at room temperature[16]. Because of strong phonon side band emission, the zero-phonon-line of $T_{V2}$ at 919 nm[8] is not observable. To enhance fluorescence emission from a single $T_{V2}$ center being especially important for spin measurements we milled solid immersion lenses (SIL) with 20 μm diameter into the sample (see Fig. 1b and SI). Figure 1c shows the confocal fluorescence microscope image of one of the fabricated SILs. We found roughly, 20 defect centers in a 20 μm diameter SIL. To prove single photon emission, we measured the photon intensity auto-correlation. At zero delay, $g^{(2)}(0)$ is found to be well below 0.5 (see Fig. 2b), a clear indication that spots seen under the SIL mark single defect emission[21]. We also compared the collection efficiency of the SILs for various emitters inside the SIL. The maximum enhancement stems from a single defect near the centroid of the sphere (Fig. 2c). Importantly, the fluorescence emission from single $T_{V2}$ centers is stable (Fig. 2d): Even 1 mW laser illumination over months did not alter the fluorescence emission rate. We did not find blinking in contrast to previous experiments on single defect centers, $C_{Si}V_C$, in SiC[21]. We also observed bunching in the measured auto-correlation function in Fig. 2b. This suggests the existence of a metastable shelving state, which is in agreement with the previously suggested model for $T_{V2}$ centers[14,16,22,23]. For room temperature spin polarization and optical detection of spin resonance the existence of such metastable states is essential as explained below.

Figure 3a shows the suggested energy-level scheme of the $T_{V2}$ center at zero magnetic field. The electronic ground state of the $T_{V2}$ center is known to have a quartet manifold (S=3/2)[22,23] and can be described by the electronic spin Hamiltonian

$$H = D\left[S_z^2 - S(S+1)/3\right] + E\left(S_x^2 - S_y^2\right) + g\mu_B \mathbf{B}_0 \cdot \mathbf{S} \qquad (1)$$

where S=3/2, g is the Landé g-factor of the ground state of $T_{V2}$ (known as $T_{V2a}$) (g=2.0028[8]), $\mu_B$ is the Bohr magneton, $\mathbf{B}_0$ is the applied static magnetic field. When there exists dipolar spin-spin interaction, spin sublevels, $m_s=|3/2|$ and $m_s=|1/2|$ states are split by the zero-field-splitting (ZFS) parameters, D and E ($ZFS = 2\sqrt{D^2 + 3E^2}$ (see SI)). D is known to be 35 MHz for $T_{V2a}$ and $E \ll D$ due to the uniaxial symmetry in the ground state[8]. The excited state lifetime is known to be 6.1 ns[24]. The intersystem-crossing (ISC) rates from and to the shelving state are not yet known precisely, but strongly depend on the spin states of the ground and excited state so that polarization into $|m_s = \pm 1/2\rangle$ ground states by light illumination[16] can be observed even at room temperature[14]. Thus, once the ground state spin polarization is altered by resonant RF irradiation, a change in the fluorescence emission intensity occurs (optically detected magnetic resonance, ODMR). Indeed, room temperature ODMR signals have been detected from $T_{V2a}$ ensembles[23].

To test whether a spin signal can be found from single defects, we first performed experiments under constant light illumination (0.1 mW). To completely lift spin state degeneracy an axial magnetic field of 50 G (see SI) is applied. Spin transitions were induced by a RF field applied by a 20 μm diameter copper wire spanned across the sample. We observed two ODMR resonance signals with narrow Lorentzian lineshapes (FWHM≈6 MHz) as seen in Fig. 3c. The ODMR signal intensity decreases at stronger laser illumination because of competing spin transitions and optical pumping (see SI). This is consistent with the expectation that $T_{V2a}$ is the electronic ground state of the defect[22]. The observed resonance frequencies are in agreement with the two allowed transitions $|m_s=+3/2\rangle\leftrightarrow|m_s=+1/2\rangle$ and $|m_s=-1/2\rangle\leftrightarrow|m_s=-3/2\rangle$ as confirmed by the spin Hamiltonian in equation (1). Calculation predicts another transition between $|m_s=+1/2\rangle\leftrightarrow|m_s=-1/2\rangle$. The absence of this resonance has been discussed previously for ensemble experiments[23]. We tentatively attribute this to identical population of the two sublevels of $|m_s=\pm1/2\rangle$. To verify the S=3/2 model of the electronic spin, experiments were repeated while changing the axial magnetic field strength, $|B_{0,z}|$ from 0 to 70 G with stronger RF radiation (power broadened FWHM≈13 MHz). The observed resonance frequencies (Fig. 3d) are in good agreement with predictions for $T_{V2a}$ using equation (1) (see Fig. 3b and SI). However, the observed signal at zero-magnetic field unveils additional fine structure. We also observed a similar structure from a $T_{V2a}$ ensemble (see SI). Such fine structure at zero field cannot be explained by our spin Hamiltonian even with the known hyperfine coupling to $^{29}Si$ and $^{13}C$ and is still under investigation.

We proceed to test the room temperature spin coherence of a single $T_{V2a}$ at $|B_{0,z}|\geq50$ G. In the following, we choose two states, $|m_s=+3/2\rangle$ and $|m_s=+1/2\rangle$ to measure coherent spin dynamics (see SI). To demonstrate the coherent manipulation of a single electron spin in SiC, we first establish optical polarization into $|m_s=+1/2\rangle$ by applying a 400 ns long laser pulse followed by a 600 ns delay to ensure that the decay from the shelving state to the GS is complete. We then apply a rectangular RF pulse resonant to the $|m_s=+3/2\rangle\leftrightarrow|m_s=+1/2\rangle$ transition, followed by an identical laser pulse for readout. This sequence (Fig. 4a) is repeated while increasing the RF pulse length up to 3 μs. The result (Fig. 4c) shows long-lived spin Rabi oscillations. The expected linear dependence of the Rabi frequency on the $B_1$ field strength is confirmed in Fig. 4b. In order to quantify electron spin coherence of a single spin in SiC, we recorded Hahn-echo decays at $B_0\approx270$ and 288 G. As before, we used a 400 ns long laser pulse for polarization and readout before and after the projective Hahn echo-sequence, respectively (Fig. 4d). An apparent feature is the strong envelope modulation (Fig. 4d) which we attribute to hyperfine coupling to a proximal $^{29}Si$ nuclear spin at close distance (e.g., ≤ ~1 nm) from the defect electron spin (see SI)[9,15,25–27]. The middle panel in Fig. 4d ($B_0$ = 288 G) shows a distinctly different modulation pattern with modulation frequencies different from the lower panel ($B_0$ = 270 G). In both case the data quality and the complex modulation pattern does not allow to determine a conclusive decoherence time. The data set in the lower panel ($B_0$= 270 G) shows also envelope modulations whose dominant modulation frequency is as similar with the theoretical prediction. While the observed curve still shows beating patterns, which is not predicted by our model and may be due to the misalignment of the applied $B_0$ field, we note that the maximum amplitude is revived at around 160 μs. Therefore, 160 μs should be considered as a lower boundary for $T_2$. The upper boundary is given by electron spin relaxation $T_1$ which we measured to be $T_1$=500 μs (see SI). Hence as upper limit we expect $T_2$ to be around 1 ms (see SI). If not limited by spin-phonon interaction, magnetic fluctuations arising from surrounding nuclear and electron spin baths are the two main sources of decoherence. Since the silicon vacancy electron spin is highly localized around the defect[8], its coupling to the bath nuclear spins is mainly of dipolar form. The coherence time due to the dipolar coupled bath spins is inversely proportional to the bath spin concentration[28]. In SiC, two types of nuclear spin species, namely, $^{29}$Si and $^{13}$C (with natural abundance 4.7 % and 1.1 %, respectively) form the nuclear spin bath. Because of higher abundance of $^{29}$Si than $^{13}$C, intuitively, one may expect faster decoherence in SiC nuclear spin bath than e.g. in diamond. However, since the gyromagnetic ratio difference between two nuclear spin types is ~150 Hz/G, the Zeeman energy difference is larger compared to the typical dipolar coupling between nuclear spins (~100 Hz) at the given magnetic field strength. Hence, nuclear spin flip-flops are greatly suppressed. As a result, $^{29}$Si and $^{13}$C nuclear spins behave as two independent spin baths. In addition, the large Si-C bond length in SiC and small gyromagnetic ratio of $^{29}$Si further decrease the effective nuclear spin bath concentration. In total, the effective nuclear spin bath concentration in SiC is quite similar to that in diamond and one would expect a couple of hundred μs and even up to an order of miliseconds for $T_2$ (see Methods and SI). Suppression of such decoherence can be achieved by diluting the remaining nuclear spin bath, in isotopically purified host crystals which has become available recently[29]. The electron spin coherence time in solids is also influenced by an electron spin bath of surrounding electron paramagnetic impurities. For example, the $T_2$ of NV centers in diamond can be significantly shortened to a few μs when the concentration of electron paramagnetic impurities is high[30]. The sample used in this study already contains quite an amount of impurities, i.e. N and B impurities up to $5\times10^{15}$ cm$^{-3}$ [31]. It also contains a variety of intrinsic point defects even before irradiation such as carbon vacancies[32], carbon vacancy-antisite pairs[21], and divacancies[9,32] which have an electronic spin of 1/2 or 1 and concentration of approximately up to ~3-6$\times10^{15}$ cm$^{-3}$ [32]. Assuming that the central spin is coupled to the bath electron spins via dipole-dipole interaction with an electron spin concentration of around $10^{16}$ cm$^{-3}$, we estimate the decoherence time in the given sample to be of the order of 100 μs (see SI). This decoherence mechanism can be suppressed by applying high magnetic fields[30] as well as dynamical decoupling[33]. Reducing the common residual B acceptor and N donor in CVD-grown layers to $10^{13}$-$10^{14}$ cm$^{-3}$ while retaining n-type conductivity, which is necessary for the formation of the negative Si vacancy, by optimizing growth condition and using modified CVD growth method is common nowadays[34,35]. The most abundant intrinsic defect in as-grown SiC CVD layers is the carbon vacancy whose concentration can also be reduced to the $10^{12}$ cm$^{-3}$ range without creating the Si vacancy in detectable level by controlling the Si/C ratio of precursor gases. Such defect reduction in CVD layers can lower the electron spin dipolar coupling strength by up to three orders of magnitudes. Thus, diluting the electron spin bath concentration is a further option.

Our results show that the SiC can host single point defects with long spin coherence times, which hold promise for long-lived qubits. This applies not only to the specific point defect used in this study, but also to other deep defect qubit candidates in SiC, like carbon vacancies, divacancies, and antisite-vacancy pairs[7] and also other point defects such as substitutional defects[7], i.e. the N impurity[10] and transition metal defects[7]. A major challenge in the present work is the weak spin signal of single $T_{V2a}$ centers. It, however, has been shown that $T_{V2a}$ center ensemble shows drastically enhanced optical spin signal at cryogenic temperature up to more than 100 % relative intensity[16], which will shorten the measurement time by roughly a factor of one hundred. This combined with the narrow PL line (<<1 nm) of $T_{V2}$ centers at low temperature [36] and mature fabrication methods, strongly suggest that SiC is a promising platform for integrating spintronics, electronics and photonics in a single quantum system.

*Note added in proof.* During the preparation of the manuscript, we became aware of a similar study[37] reporting the optical detection of single spins in SiC using the other type of defect; divacancies. This report also presents millisecond coherence times, which we theoretically predict in this study, from divacancy ensembles.

**Methods**

**Experimental setup**

For the creation of single defects, commercial on-axis high-purity semi-insulating 4H-SiC substrates were irradiated by 2 MeV electron with a fluence of $10^{13}$ - $5\times10^{14}$ cm$^{-3}$ along c-axis of the SiC crystal. The concentration of the created silicon vacancy centers is linearly dependent on the electron flux (see SI). A couple of SILs of various sizes were created on $6\times10^{13}$ cm$^{-3}$ irradiated sample by ion milling using 40 keV Ga focused ion beam. A SIL with 20 μm diameter was used in this report. A bright emission layer near the surface was found by confocal fluorescence microscope after ion milling. Optical identification of the created defect centers was only possible after this damaged layer was removed (see SI). A typical home-built confocal setup was used after optimizing for emission in the wavelength range around 900 nm. In addition, polarization optics were added to suppress the cross talk via the breakdown flash of silicon APDs, which arises at around 900 nm (see SI). The laser wavelength used to collect all of the data presented in this report was 730 nm. For spin manipulation, the RF field was irradiated via a 20 μm diameter copper wire placed over the sample surface. For continuous wave ODMR as in Fig. 3c and d, RF irradiation was applied continuously for 20 ms at each frequency and the frequency was changed in 1 MHz step. All experiments were done at room temperature. Further details about methods are found in SI.

**Estimation of the effective nuclear spin bath concentration**

The nuclear spin concentration of diamond is determined by the natural abundance of $^{13}$C (1.1 %), i.e. $\rho_{diamond} = \rho_{13}$. In SiC, taking into account two independent nuclear spin baths, the effective nuclear spin bath concentration is estimated as $\rho_{SiC} = 0.5 \times \eta_V(\rho_{13} + \rho_{29} \times \eta_\gamma)$. Here, $\rho_{29}$=4.7 % is the natural abundance of $^{29}$Si. The ratio $\eta_V = \left(\frac{d_{C-C}}{d_{Si-C}}\right)^3 = 0.55$ describes the unit cell volume expansion of SiC comparing with diamond, with $d_{Si-C}$=1.88 Å and $d_{C-C}$=1.54 Å are the Si-C and C-C bond lengths in SiC and diamond, respectively. Smaller gyromagnetic ratio of $^{29}$Si ($\gamma_{29}$=2π×0.85 kHz/G) compared to that of $^{13}$C ($\gamma_{13}$=2π×1.07 kHz/G) also reduce the nuclear spin bath flipping rate of the $^{29}$Si bath. The ratio $\eta_\gamma = \left(\frac{\gamma_{29}}{\gamma_{13}}\right)^2 = 0.63$ accounts for the smaller nuclear spin flip-flop rate between $^{29}$Si nuclei. With these numbers, the effective nuclear spin concentration of SiC is similar to that of diamond, i.e. $\frac{\rho_{SiC}}{\rho_{diamond}} \approx 1$. Consequently, SiC nuclear spin bath should cause similar electron spin decoherence rate as $^{13}$C in diamond. The estimation is consistent with more detailed numerical calculations based on the cluster expansion method which even predicts a very long coherence time up to a milliseconds (see SI)[27].

**References**

1. Morton, J. J. L. & Lovett, B. W. Hybrid Solid-State Qubits: The Powerful Role of Electron Spins. *Annu. Rev. Condens. Matter Phys.* **2,** 189–212 (2011).

2. Awschalom, D. D., Bassett, L. C., Dzurak, A. S., Hu, E. L. & Petta, J. R. Quantum Spintronics: Engineering and Manipulating Atom-Like Spins in Semiconductors. *Science* **339,** 1174–1179 (2013).

3. Balasubramanian, G. *et al.* Ultralong spin coherence time in isotopically engineered diamond. *Nat Mater* **8,** 383–387 (2009).

4. Lee, S.-Y. *et al.* Readout and control of a single nuclear spin with a meta-stable electron spin ancilla. *Nat Nano* **8,** 487–492 (2013).

5. Weber, J. R. *et al.* Quantum computing with defects. *Proc. Natl. Acad. Sci.* **107,** 8513 (2010).

6. Morello, A. *et al.* Single-shot readout of an electron spin in silicon. *Nature* **467,** 687–691 (2010).

7. Janzén, E. *et al.* in *Defects Microelectron. Mater. Devices* 615–669 (CRC Press, 2008). doi:doi:10.1201/9781420043778.ch21

8. Janzén, E. *et al.* The silicon vacancy in SiC. *Phys. B Condens. Matter* **404,** 4354–4358 (2009).

9. Koehl, W. F., Buckley, B. B., Heremans, F. J., Calusine, G. & Awschalom, D. D. Room temperature coherent control of defect spin qubits in silicon carbide. *Nature* **479,** 84–87 (2011).

10. Aichinger, T., Lenahan, P. M., Tuttle, B. R. & Peters, D. A nitrogen-related deep level defect in ion implanted 4H-SiC pn junctions---A spin dependent recombination study. *Appl. Phys. Lett.* **100,** 112113–112114 (2012).

11. Maboudian, R., Carraro, C., Senesky, D. G. & Roper, C. S. Advances in silicon carbide science and technology at the micro- and nanoscales. *J. Vac. Sci. Technol. A Vacuum, Surfaces, Film.* **31,** 50805–50818 (2013).

12. Song, B.-S., Yamada, S., Asano, T. & Noda, S. Demonstration of two-dimensional photonic crystals based on silicon carbide. *Opt. Express* **19,** 11084–11089 (2011).

13. Matsunami, H. Current SiC technology for power electronic devices beyond Si. *Microelectron. Eng.* **83,** 2–4 (2006).

14. Soltamov, V. A., Soltamova, A. A., Baranov, P. G. & Proskuryakov, I. I. Room Temperature Coherent Spin Alignment of Silicon Vacancies in 4H- and 6H-SiC. *Phys. Rev. Lett.* **108,** 226402 (2012).

15. Gaebel, T. *et al.* Room-temperature coherent coupling of single spins in diamond. *Nat Phys* **2,** 408–413 (2006).

16. Baranov, P. G. *et al.* Silicon vacancy in SiC as a promising quantum system for single-defect and single-photon spectroscopy. *Phys. Rev. B* **83,** 125203 (2011).

17. Falk, A. L. *et al.* Polytype control of spin qubits in silicon carbide. *Nat Commun* **4,** 1819 (2013).

18. Falk, A. L. *et al.* Electrically and Mechanically Tunable Electron Spins in Silicon Carbide Color Centers. *Phys. Rev. Lett.* **112,** 187601 (2014).

19. Yang, Y. T. *et al.* Monocrystalline silicon carbide nanoelectromechanical systems. *Appl. Phys. Lett.* **78,** (2001).

20. Klimov, P. V., Falk, A. L., Buckley, B. B. & Awschalom, D. D. Electrically Driven Spin Resonance in Silicon Carbide Color Centers. *Phys. Rev. Lett.* **112,** 87601 (2014).

21. Castelletto, S. *et al.* A silicon carbide room-temperature single-photon source. *Nat Mater* **13,** 151–156 (2014).

22. Mizuochi, N. *et al.* Continuous-wave and pulsed EPR study of the negatively charged silicon vacancy with S=3/2 and C_{3v} symmetry in n-type 4H-SiC. *Phys. Rev. B* **66,** 235202 (2002).

23. Kraus, H. *et al.* Room-temperature quantum microwave emitters based on spin defects in silicon carbide. *Nat Phys* **10,** 157–162 (2014).

24. Hain, T. C. *et al.* Excitation and recombination dynamics of vacancy-related spin centers in silicon carbide. *J. Appl. Phys.* **115,** 133508 (2014).

25. Childress, L. *et al.* Coherent Dynamics of Coupled Electron and Nuclear Spin Qubits in Diamond. *Science* **314,** 281–285 (2006).

26. De Sousa, R. & Das Sarma, S. Theory of nuclear-induced spectral diffusion: Spin decoherence of phosphorus donors in Si and GaAs quantum dots. *Phys. Rev. B* **68,** 115322 (2003).

27. Yang, L.-P. *et al.* Electron Spin Decoherence in Silicon Carbide Nuclear Spin Bath. *Arxiv Prepr. arXiv1409.4646* (2014).

28. Witzel, W. M., Carroll, M. S., Morello, A., Cywiński, Ł. & Das Sarma, S. Electron Spin Decoherence in Isotope-Enriched Silicon. *Phys. Rev. Lett.* **105,** 187602 (2010).

29. Ivanov, I. G. *et al.* High-Resolution Raman and Luminescence Spectroscopy of Isotope-Pure 28Si12C, Natural and 13C–Enriched 4H-SiC. in *Mater. Sci. Forum* **778,** 471–474 (Trans Tech Publ, 2014).

30. Hanson, R., Dobrovitski, V. V, Feiguin, A. E., Gywat, O. & Awschalom, D. D. Coherent Dynamics of a Single Spin Interacting with an Adjustable Spin Bath. *Science* **320,** 352–355 (2008).

31. Jenny, J. R. *et al.* Development of Large Diameter High-Purity Semi-Insulating 4H-SiC Wafers for Microwave Devices. in *Mater. Sci. Forum* **457,** 35–40 (Trans Tech Publ, 2004).

32. Son, N. T., Carlsson, P., ul Hassan, J., Magnusson, B. & Janzén, E. Defects and carrier compensation in semi-insulating 4H-SiC substrates. *Phys. Rev. B* **75,** 155204 (2007).

33. De Lange, G., Wang, Z. H., Ristè, D., Dobrovitski, V. V & Hanson, R. Universal Dynamical Decoupling of a Single Solid-State Spin from a Spin Bath. *Science* **330,** 60–63 (2010).

34. Kimoto, T., Nakazawa, S., Hashimoto, K. & Matsunami, H. Reduction of doping and trap concentrations in 4H–SiC epitaxial layers grown by chemical vapor deposition. *Appl. Phys. Lett.* **79,** (2001).

35. Tsuchida, H. *et al.* Characterization of 4H-SiC epilayers grown at a high deposition rate. in *Mater. Sci. Forum* **353,** 131–134 (Trans Tech Publ, 2001).

36. Riedel, D. *et al.* Resonant Addressing and Manipulation of Silicon Vacancy Qubits in Silicon Carbide. *Phys. Rev. Lett.* **109,** 226402 (2012).

37. Christle, D. J. *et al.* Isolated electron spins in silicon carbide with millisecond-coherence times. *arXiv Prepr. arXiv1406.7325* (2014).

**Acknowledgements**

We acknowledge support by EU via SQUTEC, and SIQS as well as QINVC. The DARPA via QuASAR. The DFG via SPP 1601 and Forschergruppe FOR1493 and the Max Planck Society. A.G. acknowledges the support from the Lendület programme of the Hungarian Academy of Sciences, and the Hungarian OTKA grants No. K101819 and K106114. Support from Knut & Alice Wallenberg Foundation (for N.T.S, A.G. and E.J.), Linköping Linnaeus Initiative for Novel Functionalized Materials (for N.T.S), and the Ministry of Education, Science, Sports and Culture in Japan, Grant-in-Aid for Scientific Research (B) 26286047 (for T.O.) is acknowledged. N.Z. acknowledges NKBRP (973 Program) 2014CB848700, and NSFC No. 11121403.. We thank Roman Kolesov, Rainer Stöhr, Phil Hemmer, Norikazu Mizuochi, and Achim Güth for fruitful discussion and experimental aid.

**Author contributions**

M.W., S.L, N.T.S., H.F., S.P, and J.W. conceived and designed the experiment; M.W., S.L., T.R., and S.P. performed the experiment; M.W., S.L., and T.R., analyzed the data; N.T.S., I.B., T.O., and E.J. contributed materials and electron irradiation; M.W., S.L., N.T.S., S.Y., I.B., A.D., M.J., S.M., and I.G. contributed to fabrication of SILs; L.Y., N.Z. and A.G. provided theoretical support; M.W., S.L., T.R., N.T.S., H.F., S.P., L.Y., N.Z., A.G., E.J., and J.W. discussed and wrote the paper.

**Competing financial interests**

The authors declare no competing financial interests.

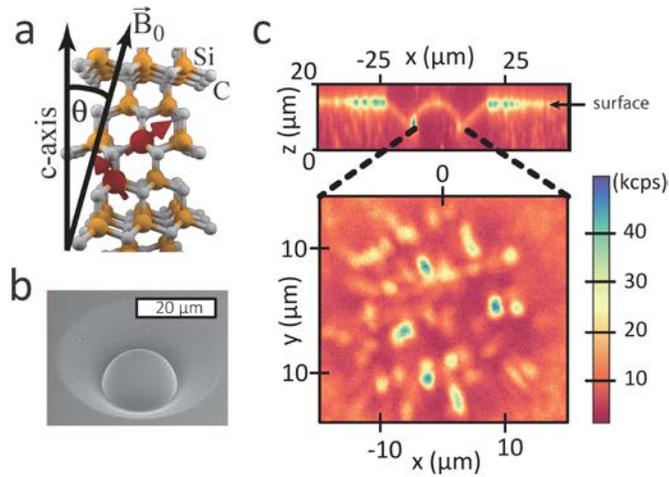

**Figure 1**: Observation of single silicon vacancy defects in 4H-SiC. **a,** 4H-SiC crystal structure and spins at silicon vacancies together with the c-axis orientation. The angle between the applied $B_0$ field orientation and the c-axis is $\theta$. **b,** SEM picture of the fabricated SIL on the surface. **c,** Confocal fluorescence image scanned around the fabricated hemispherical SIL with 12 mW at 730 nm laser excitation. Top: x-z scan showing the cross-section of the fabricated 20 μm diameter SIL. Bottom: x-y scan showing defects (bright spots) inside the SIL.

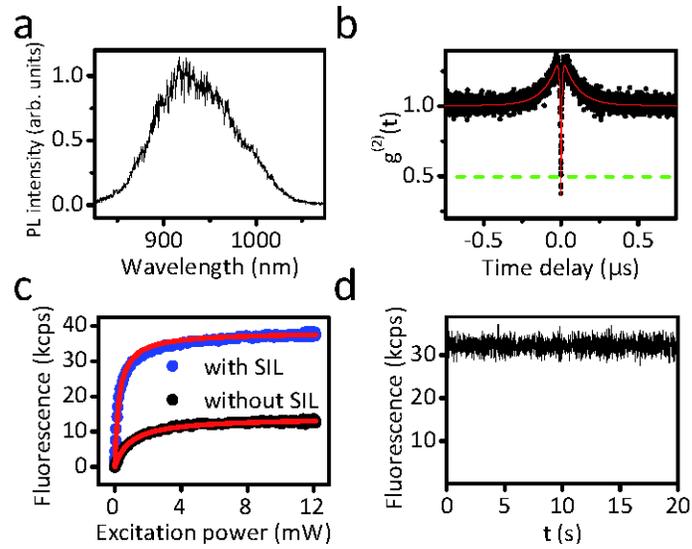

**Figure 2:** Optical properties of the single $T_{V2}$ centers in 4H-SiC. **a,** Fluorescence emission spectrum of a single $T_{V2}$ center with 1 mW 730 nm laser excitation and 805 nm LP filter. **b,** Autocorrelation measurement of a single $T_{V2}$ center measured at 0.1 mW optical power. Experimental data are plotted after background correction and deconvoluting timing jitter of APDs ($\approx$1.39 ns). Red curve is a fit based on the three states model (Fig. 3a) and green line indicates $g^{(2)}$=0. **c,** Saturation curves of single $T_{V2}$ centers with and without a SIL. The saturation intensity is increased by a factor of up to 3. Red curves: fits based on the three states model (Fig. 3a). **d,** Time trace of fluorescence intensity of a single $T_{V2}$ center with 10 ms bins showing stable fluorescence emission from a single $T_{V2}$ center.

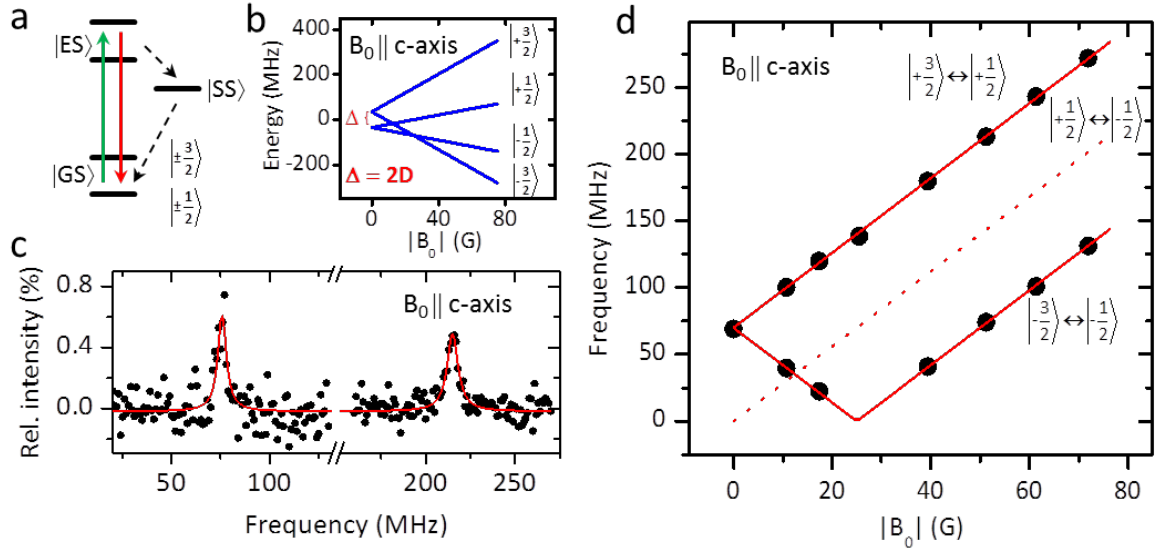

**Figure 3:** Optically detected electron spin resonance of a single $T_{V2}$ center at room temperature. **a,** Suggested energy-level diagram showing the ground (GS), excited (ES), and shelving state (SS). Solid lines: Radiative transition (red) and optical pumping (green). Dashed lines: non-radiative transitions, presumably spin-dependent, resulting in GS spin polarization. For the fit functions in Fig. 2b and d, ES and GS are assumed to consist of a single state for simplicity. **b,** Calculated energy eigenvalues, expressed in the unit of frequency, of each spin sublevel as a function of the axial $B_0$ field strength. **c,** ODMR spectrum of a single $T_{V2a}$ center at $|B_0|$=50 G with $B_0 \parallel$ c-axis ($\theta=0\pm4°$). Black dots: measured data expressed as a relative fluorescent intensity (= $\Delta PL/PL_{off}$, $PL_{off}$: PL intensity at off-resonance). Red curves: Lorentzian fit with full width at half maximum (FWHM) $\approx$6 MHz. **d,** Experimentally obtained frequency dependence of ODMR lines with axial $B_0$. Red curves are based on calculations for the allowed transitions using Equation (1). Dashed red line: expected $|m_s=-1/2\rangle\leftrightarrow|m_s=-3/2\rangle$ transition (see text).

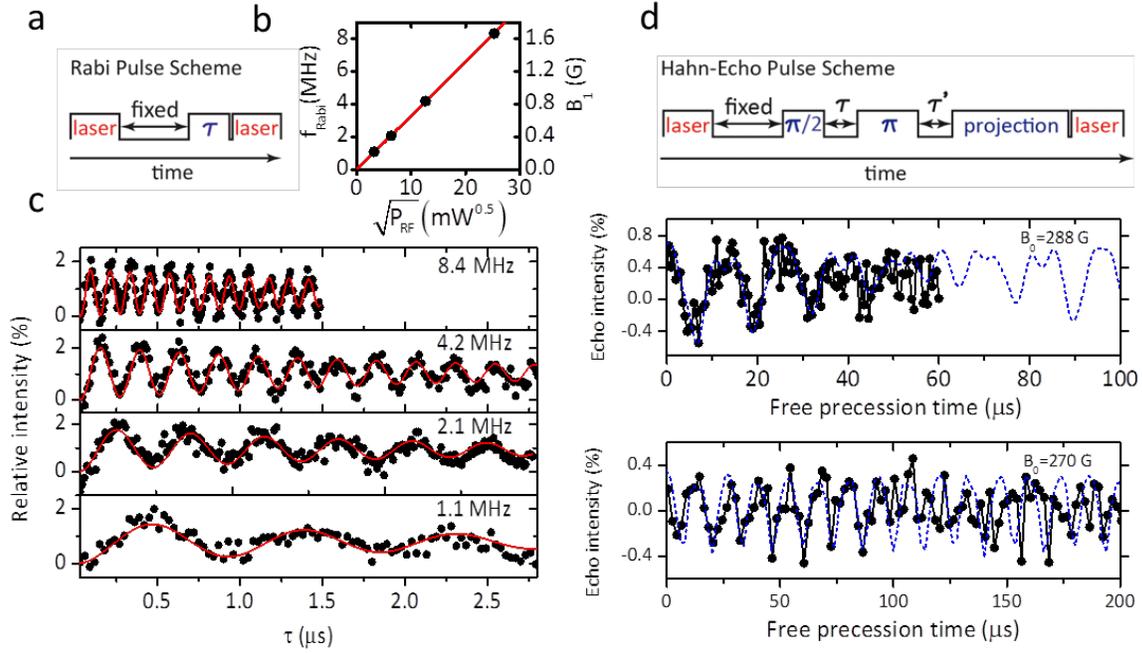

**Figure 4:** Room temperature coherent spin manipulation of a single $T_{V2}$ center in 4H-SiC. **a,** Pulse sequence for detecting spin Rabi oscillations (see text). **b,** Measured Rabi frequencies ($f_{Rabi}$), extracted from (c), and converted RF field ($B_1$) strength vs the square root of the total RF power ($P_{RF}$) applied to the wire. For conversion, $f_{Rabi} = \sqrt{3} g \mu_B B_1 / h$ is used[22]. Red line is a linear fit. **c,** Spin Rabi oscillations of the $|m_s = +3/2\rangle \leftrightarrow |m_s = +1/2\rangle$ transition of a single $T_{V2a}$ center at $|g\mu_B B_{0,z}| \approx 4D \approx 50$ G and various $B_1$ field strength. The spin signal is obtained by integrating the photon counts in the first 60~120 ns of the fluorescence response to the readout laser pulse. Red curves: exponentially decaying sinusoidal fits. The numbers indicate the extracted Rabi frequencies, which are linearly dependent on the $B_1$ field strength as shown in (b). **d,** Hahn echo decay at $|B_{0,z}| \approx 270$ and 288 G. Top: pulse sequence consisting of $\pi/2$ and pi pulses separated by $\tau$, and an additional $\pi/2$ for projection after another delay $\tau'$. The same sequence is repeated by replacing the last $\pi/2$ pulse with $3\pi/2$ pulse. The difference between the two data, normalized by the average of them is plotted as an echo intensity to remove any contribution from unknown relaxation processes. Middle and bottom: the measured Hahn-echo at 288 and 270 G, respectively. The blue curve shows the simulated Hahn-echo modulated by the coupling to a proximal $^{29}$Si nuclear spin (see SI).

# Supplementary information for "Coherent control of single spins in silicon carbide at room temperature"

Matthias Widmann[1], Sang-Yun Lee[1*], Torsten Rendler[1], Nguyen Tien Son[2], Helmut Fedder[1], Seoyoung Paik[1], Li-Ping Yang[3], Nan Zhao[3], Sen Yang[1], Ian Booker[2], Andrej Denisenko[1], Mohammad Jamali[1], Seyed Ali Momenzadeh[1], Ilja Gerhardt[1], Takeshi Ohshima[4], Adam Gali[5,6], Erik Janzén[2], Jörg Wrachtrup[1]

1. *3rd Institute of Physics and Research Center SCOPE, University of Stuttgart, Pfaffenwaldring 57, 70569 Stuttgart, Germany*
2. *Department of Physics, Chemistry and Biology, Linköping University, SE-58183 Linköping, Sweden*
3. *Beijing Computational Science Research Center, Beijing 100084, China*
4. *Japan Atomic Energy Agency, Takasaki, Gunma 370-1292, Japan*
5. *Wigner Research Centre for Physics, Hungarian Academy of Sciences, P.O. Box 49, H-1525, Budapest, Hungary*
6. *Department of Atomic Physics, Budapest University of Technology and Economics, Budafokiút 8, H-1111, Budapest, Hungary*

*e-mail: s.lee@physik.uni-stuttgart.de

## 1. Experimental Setup

A home-built confocal microscope was used to perform measurements (see Fig. S1). A red diode laser (Thorlabs HL7302MG, emission peak wavelength 730 nm, 40mW) was focused via an oil objective (Olympus, UPLSAPO 60XO, NA 1.35) onto the sample to excite the silicon vacancies ($T_{V2}$ centers). With 730 nm excitation wavelength the 1st and 2nd order Raman emissions of SiC[1] and Raman emissions of the used immersion oil (Fluka 10976) show up in the wavelength range up to 900 nm. Hence a 905 nm long-pass filter was installed in front of the photon counting detectors. Another emission band from the immersion oil still appears at around 930 nm but the contribution is weak. The photoluminescence (PL) spectrum presented in the main text (Fig.2a) was measured with a 805 nm long-pass filter and is corrected for the remaining background. The incident laser beam was parallel to [0001] orientation of the used SiC crystal (c-axis) and linearly polarized. The $T_{V2}$ center has a single optical dipole transition with a dipole axis oriented parallel to the defect symmetry axis[2]. Without focusing, no optical dipole excitation would be possible in this configuration. However, consider the marginal rays passing through the high N.A. microscope objective. For these rays, the electric field vector has a significant projection onto the dipole axis, providing efficient dipole excitation. PL spectra were recorded using a Czerny–Turner type spectrometer (Acton, SpectraPro300i, grating: 300 g mm$^{-1}$) combined with a back-illuminated CCD camera (Princeton Instruments, LN/CCD-1340/400 EHRB/I) with a 800-nm long pass filter (Thorlabs FELH0800). Two single photon counting avalanche photodiodes (APD) (PerkinElmer SPCM-AQRH-15) in Hanburry-Brown and Twiss configuration (HBT)[3] were used for detection and to measure photon intensity auto-correlation. To suppress APD cross talk due to breakdown flashes generated in the APDs[4], a polarizing beam splitter (PBS, Thorlabs CM1-PBS252) was used in combination with Glan-Taylor calcite polarizers (GTP, Thorlabs GT10-B) in front of each APD (see Fig. S1). In addition, an achromatic half-wave plate (Thorlabs AHWP10M-980) was placed in front of the PBS to equilibrate the count rates on both detectors.

An electro-magnet mounted to a motorized 2D positioning system (Standa 8MTF), was used to provide the static magnetic field parallel to the $T_{V2}$ center axis (c-axis). Here, a field strength up to 290 G was applied. Electron spins were manipulated by a radio-frequency electromagnetic field generated by a signal generator (ROHDE&SCHWARZ, SMIQ03B) and subsequently amplified by a broadband amplifier (Minicircuits ZHL-42W). RF irradiation was applied via a 20 μm thick copper wire, which is placed close to the tested defects (typically ~30 μm). For other details, see refs [5,6].

## 2. Defect Concentration

Silicon vacancies were created by 2 MeV electron irradiation on a 510 μm thick commercial on-axis high-purity semi-insulating 4H-SiC single crystal sample. The electrons penetrate the whole sample, creating defects homogeneously through the whole volume. During irradiation, water cooling kept the sample temperature around room temperature. To suppress the creation of other types of defects, no subsequent annealing was performed[7]. The defect concentration decreases with decreasing radiation dose as shown in Fig. S2. Electron fluences below $6\times10^{13}$ cm$^{-2}$ allowed us to resolve individual defects. Once the average total intensity from a single silicon vacancy center was determined at a given laser power, the total number of created defect centers were counted by confocal fluorescence scanning over a large area, ≈100 um × 100 um. The result is described in Fig. S2, showing an approximately linear dependence between the concentration of the silicon vacancy centers and the used electron fluence.

## 3. Fabrication of solid immersion lenses

All measurements of the electron spin state of the silicon vacancy centers were based on detection of the $T_{V2}$ center PL in a bulk SiC crystal. Most of the $T_{V2}$ center PL, however, is lost at the crystal surface due to the total internal reflection. Hence, efficient photon detection is necessary. We enhance the photon collection efficiency by fabricating a solid immersion lens (SIL), into the polished surface of the silicon carbide crystal by focused ion beam milling (FIB) (FEI Helios Nanolab 600, 40 keV)[8], as shown in Fig. S3. To prevent surface charging during the milling process, a 50 nm gold coating is sputtered onto the surface prior to the FIB process and removed afterwards by aqua regis. After this process, a bright fluorescing surface layer was still observed in the confocal microscope as shown in Fig. S3a. This layer may contain various defects created during the ion milling. It results in a prohibitive fluorescence background that must be removed. At the same time, the ion irradiation also creates an amorphous SiC layer[9] which acts as an etch-stop for conventional plasma etching. Thus, the fluorescing layer was removed by first wet etching the amorphous layer by placing the sample in a mixture of 49% HF + 69% $HNO_3$ (1:1) for 4 hours and subsequently etching the surface by about 200 nm using inductively coupled plasma (ICP) etching. After the ICP etching step, the removal of the fluorescent layer was confirmed by confocal microscopy (Fig. S3b).

**4. Spin Hamiltonian for the ground state of $T_{V2}$ centers**

The observed ODMR data of $T_{V2a}$ centers can be described with a spin Hamiltonian given by

$$H = D\left[S_z^2 - \frac{S(S+1)}{3}\right] + E\left(S_x^2 - S_y^2\right) + g\mu_B \mathbf{B}_0 \cdot \mathbf{S}$$

Where $g$ is the isotropic Landé $g$ factor (g=2.0028), $\mu_B$ is the Bohr magneton, S=3/2, and $\mathbf{B}_0$ denotes the externally applied magnetic field. The dipolar magnetic interaction is described by the zero-field-splitting (ZFS) parameters, D and E: D is known to be 35 MHz and positive for $T_{V2a}$ and E<<D due to the uniaxial symmetry in the ground state[10,11]. The matrix form of the above Hamiltonian in the |S,m$_s$⟩ basis, where m$_s$=+3/2,+1/2, -1/2, and -3/2, is

$$\begin{pmatrix} \frac{3g\mu_B B_0}{2} + D & 0 & \sqrt{3}E & 0 \\ 0 & \frac{3g\mu_B B_0}{2} - D & 0 & \sqrt{3}E \\ \sqrt{3}E & 0 & -\frac{3g\mu_B B_0}{2} - D & 0 \\ 0 & \sqrt{3}E & 0 & -\frac{3g\mu_B B_0}{2} + D \end{pmatrix}$$

for $\mathbf{B}_0$ applied parallel to the spin quantization axis (c-axis), and the eigenvalues of this spin Hamiltonian with corresponding eigenstates are

$$\frac{g\mu_B B_0}{2} + \sqrt{(g\mu_B B_0 + D)^2 + 3E^2} \text{ for } |m_s = 3/2\rangle,$$

$$-\frac{g\mu_B B_0}{2} + \sqrt{(g\mu_B B_0 - D)^2 + 3E^2} \text{ for } |m_s = 1/2\rangle,$$

$$\frac{g\mu_B B_0}{2} - \sqrt{(g\mu_B B_0 + D)^2 + 3E^2} \text{ for } |m_s = -1/2\rangle,$$

$$-\frac{g\mu_B B_0}{2} - \sqrt{(g\mu_B B_0 - D)^2 + 3E^2} \text{ for } |m_s = -3/2\rangle,$$

These are plotted in Fig. S4a for D= 35 MHz and E<<D. At zero magnetic field (B₀=0), the splitting between, |m_s|=3/2 and |m_s|=1/2 states is $2\sqrt{D^2 + 3E^2}$ (≈2D for E<<D), thus the deviation from the uniaxial symmetry, resulting in non-zero E value, does not eliminate the remaining spin state degeneracy of T$_{v2a}$ at zero magnetic field. Note that because S=1 was assigned to T$_{V2}$ centers at the early stage studies, D was reported to be 70 MHz in some publications[10,12].

**5. B₀ field alignment**

In order to align the magnetic field along the spin quantization axis, c-axis, an electro-magnet was mounted over the sample on an x-y stage. According to the simulation in Fig.S4b showing the resonant frequencies of the allowed transitions for a fixed high B₀ field (50 G) as a function of polar angles (see Fig.1a in the main text), the splitting between |+3/2⟩↔|+1/2⟩ and |-3/2⟩↔|-1/2⟩ transitions reaches a maximum of 4D, when the **B₀** is aligned along c-axis. The B₀ field strength should be larger than 2D in order to make the splitting independent of |B₀|. Note that the dependence on the azimuthal angle is ignorable due to the uniaxial symmetry. For the alignment, the electromagnet was scanned over the sample while monitoring the splitting between two outer transitions until the maximum value was achieved. The applied magnetic field strength was calculated using the above equations.

## 6. Spin signal strength as a function of optical excitation power

Figure S5 shows the dependence of the ODMR signal on the optical excitation power under continuous optical excitation and RF irradiation. This behavior is consistent with a rate model of the optical cycle shown in Fig. S5c. To capture the essence of the optical cycle we consider only two of the quartet spin states, denoted 'up' and 'down' and neglect the other two. We expect that the ODMR originates from the electronic ground state, since the observed spin coherence time of ~100 µs is rather long. If the ODMR originates from a metastable excited state, the optical lifetime of such a state needed to be at least 100 µs, which is unlikely. We describe the metastable state by a single level. The symbols $\gamma_{ex}$, $\gamma_{em}$ are the optical excitation and decay rate, $\gamma_{pup}$, $\gamma_{pdown}$, $\gamma_{dup}$, $\gamma_{ddown}$, are the population- and decay rates of the shelving states for the spin up and down manifolds, $\gamma_m$ is the microwave transition rate. From the solutions to the rate equations, we find that with this model, a positive ODMR contrast and the observed dependence on the optical excitation power is observed. For this we assume that the overall decay rate of the metastable state $\gamma_d = \gamma_{dup} + \gamma_{ddown}$ is larger than the emission rate and for one of the spin manifolds both the decay and population rate are smaller than the rates of the other spin manifold, i.e., $\gamma_{dup} < \gamma_{ddown}$ implies $\gamma_{pup} < \gamma_{pdown}$. Such a rate model consistently describes the data. It should be noted that, with a suitable combination of rates, other, more complex rate models will also describe the observed ODMR contrast.

**7. Fine structure at zero magnetic field**

Figure S6 shows the zero-field ODMR spectra of a chosen single $T_{V2a}$ center and a $T_{V2a}$ ensemble. This particular single defect center was used for the study presented in the main text. Sharp peaks are found in the ODMR spectrum at 51, 65, 76, and 90 MHz from a single $T_{V2a}$ center (Fig. S6a). Similarly, multiple peaks are also found at similar frequencies, 51, 54, 77, and 90 MHz except for another small peak at 43 MHz from a $T_{V2a}$ ensemble (Fig. S6b). These peaks do not originate from hyperfine (HF) interaction with a proximal nuclear spin, as we will describe in detail. Note also that we carefully checked that the peaks are not artefacts due to power dependent heating of the microwave wire, as we carefully checked by varying the driving microwave power. In natural abundant SiC, two nuclear spins are frequently observed: $^{29}$Si and $^{13}$C with natural abundance of 4.7 % and 1.1 %, respectively. The multi-peak structure cannot be due to a single $^{13}$C in the nearest-neighbor position (NN), because in this case the HF coupling strength is much larger (>30 MHz)[13] than the observed splitting between each peak (11~13 MHz). The probability for one $^{29}$Si atom to be found at one of the 12 next-nearest-neighbor (NNN) positions is 33%, much higher than 4.2 % of $^{13}$C in the nearest-neighbor (NN) position[13,14]. However, because a $^{29}$Si atom has I=1/2 and the HF coupling strength, which is isotropic, is 8.7 MHz at the NNN position[13], the observed multiple peak structure split by 11~13

MHz cannot be explained by a NNN $^{29}$Si nuclear spin. In addition, when an axial $B_0$ field is applied no such splitting was observed as can be seen in Fig. 3 in the main text. Thus we conclude that proximal nuclear spins are not responsible for the observed multiple peak structure at zero magnetic field.

The excited state ESR transitions of $T_{V2}$ centers (called $T_{V2b}$) may overlap with $T_{V2a}$ signals. However, the spin signal of $T_{V2b}$ centers can be easily distinguished because of its weaker ZFS, 2D=36 MHz[15,16].

## 8. Measurement and analysis of $T_1$

In order to measure the $T_1$ relaxation between in the ground state at $B_0$=288 G, the 400 ns laser pulse is used for the optical polarization and the same length laser pulse is applied after a delay t. Then again, the very beginning part of the fluorescence response to the readout pulse is integrated ($I_0(t)$). In order to remove any unknown relaxation processes, again the same sequence is applied with a RF $\pi$ pulse in front of the readout pulse, and the fluorescence response is integrated ($I_\pi(t)$). The difference between two intensities, $I_\pi(t) - I_0(t)$, is then normalized by the average of two values, $(I_\pi(t) + I_0(t))/2$ and plotted in Fig.S7.

In order to extract the $T_1$ from the obtained curve, we should derive a formulism for $T_1$ in this experimental configuration. We assume that the 400 ns longer laser pulse is long enough to induce the steady state at the ground state and we ignore the relaxations from the metastable states to the ground states which are very fast compared to $T_1$ so that they are not detectable in our experiments. If a very strong polarization is supposed, which is realistic because more than 80 % polarization has been reported[12], we can assume that all spin populations can be found only in $m_s=\pm 1/2$ states of the ground state. We also assume that $m_s=\pm 1/2$ states are equally populated because the transition between these two states has not been found by ODMR[15]. The relaxation is then described by the rate $T_1^{-1}$ as shown in Fig.S7. Then the rate equations for the populations of four sublevels of the ground state become

$$\frac{dn_{+3/2}}{dt} = \frac{n_{+1/2} - n_{+3/2}}{T_1}$$

$$\frac{dn_{+1/2}}{dt} = \frac{n_{+3/2} - n_{+1/2}}{T_1} + \frac{n_{-1/2} - n_{+1/2}}{T_1}$$

$$\frac{dn_{-1/2}}{dt} = \frac{n_{+1/2} - n_{-1/2}}{T_1} + \frac{n_{-3/2} - n_{-1/2}}{T_1}$$

$$\frac{dn_{-3/2}}{dt} = \frac{n_{-1/2} - n_{-3/2}}{T_1}$$

Using the initial conditions $n_{+3/2} = n_{-3/2} = 0$ and $n_{+1/2} = n_{-1/2} = 1/2$, we get the solutions

$$n_{\pm 3/2} = \frac{1 - \exp(-2t/T_1)}{4}$$

$$n_{\pm 1/2} = \frac{1 + \exp(-2t/T_1)}{4}$$

Once the readout laser pulse is applied, the sublevels of the excited states will keep the same populations as they have in the ground state if these transitions are spin-conservative. Some of these excited states, however, will experience very fast decays to the metastable states while the other sublevels relative quickly decay to the ground state and emit photons. These spin-dependent PL emission rates can be described by $R_i$ (unknown) where i=±3/2, ±1/2. Then when no π pulse is applied, the obtained PL intensity $I_0(t)$ is given by

$$I_0(t) = \sum R_i \times n_i$$

while

$$I_\pi(t) = R_{+3/2} \times n_{+1/2} + R_{+1/2} \times n_{+3/2} + \sum_{i=-1/2,-3/2} R_i \times n_i .$$

Therefore, the difference between two values is

$$I_\pi(t) - I_0(t) = \frac{R_{+3/2} - R_{+1/2}}{2} exp(-2t/T_1) .$$

Note that the time constant is not $T_1$ but $T_1/2$. Using this solution, we obtain $T_1$=500±160 μs from the data in Fig.S7.

### 9. Numerical calculations of decoherence time and electron spin echo envelope modulation

We numerically simulate the spin decoherence of the $T_{V2}$ electron spin in SiC nuclear spin bath. Similar to the NV center in diamond[17,18], the nuclear spin bath induced decoherence is considered with a pure dephasing model[19], and cluster-expansion are performed to calculate the coherence time. As shown in Ref.[19], the $T_2$ time of $T_{V2}$ centers is longer than the NV centers, and reach more than 1 ms. Figure S8 shows the calculated Hahn echo coherence of a $T_{V2}$ center. Strong coherence modulations appear on a smooth overall decay. A close-up of the modulations is presented in the inset of Fig. S8, which is identical to the blue curve in Fig.4 in the main text. The modulations are caused by the coupling to a proximal $^{29}$Si nuclear spin around 5 Å which result in multiple modulation frequencies due to the fact that all the sublevels of the spin quartet state have non-zero spin number in contrast to the NV centers in diamond and divacancies in SiC which have S=1.

### 10. Estimation of the dipolar electron spin-spin interaction
The electron spin dipole-dipole interaction between two spin, $S_i$ and $S_j$, at $r_i$ and $r_j$, respectively, is described by the Hamiltonian

$$H_{d,ij} = \frac{\mu_0 g_i g_j \mu_B^2}{4\pi \hbar r_{ij}^3} \left( \mathbf{S}_i \cdot \mathbf{S}_j - 3\frac{(\mathbf{S}_i \cdot \mathbf{r}_i)(\mathbf{S}_j \cdot \mathbf{r}_j)}{r_{ij}^2} \right)$$

where $g_i$ is the Landé g-factor of each spin species, $r_{ij}$ is the distance between them, and $\mu_0$ is vacuum permeability. Then the coupling strength of the electron spin bath is approximately determined by $\mu_0 g_e^2 \mu_B^2 / 4\pi \hbar \langle r \rangle^3$ where $g_e$ is the free electron Landé g-factor. When the paramagnetic electron spin concentration of ~$10^{16}$ cm$^{-3}$ (equivalently, average separation $\langle r \rangle$ ~20 nm[20]) is assumed, we obtain the dipolar interaction strength has the order of 10 kHz.

**References**

1. Burton, J. C., Sun, L., Long, F. H., Feng, Z. C. & Ferguson, I. T. First- and second-order Raman scattering from semi-insulating 4H-SiC. *Phys. Rev. B* **59,** 7282–7284 (1999).

2. Janzén, E. *et al.* The silicon vacancy in SiC. *Phys. B Condens. Matter* **404,** 4354–4358 (2009).

3. Brown, R. H. & Twiss, R. Q. LXXIV. A new type of interferometer for use in radio astronomy. *Philos. Mag.* **45,** 663–682 (1954).

4. Kurtsiefer, C., Zarda, P., Mayer, S. & Weinfurter, H. The breakdown flash of silicon avalanche photodiodes-back door for eavesdropper attacks? *J. Mod. Opt.* **48,** 2039–2047 (2001).

5. Balasubramanian, G. *et al.* Ultralong spin coherence time in isotopically engineered diamond. *Nat Mater* **8,** 383–387 (2009).

6. Zhao, N. *et al.* Sensing single remote nuclear spins. *Nat Nano* **7,** 657–662 (2012).

7. Baranov, P. G. *et al.* Silicon vacancy in SiC as a promising quantum system for single-defect and single-photon spectroscopy. *Phys. Rev. B* **83,** 125203 (2011).

8. Jamali, M. *et al.* Fabrication of Solid-Immersion-Lenses by focussed ion beam milling. *arXiv Prepr. arXiv1406.7209* (2014).

9. Menzel, R., Bachmann, T., Wesch, W. & Hobert, H. Maskless sub-mu m patterning of silicon carbide using a focused ion beam in combination with wet chemical etching. *J. Vac. Sci. Technol. B Microelectron. Nanom. Struct.* **16,** 540–543 (1998).

10. Sörman, E. *et al.* Silicon vacancy related defect in 4H and 6H SiC. *Phys. Rev. B* **61,** 2613 (2000).

11. Orlinski, S. B., Schmidt, J., Mokhov, E. N. & Baranov, P. G. Silicon and carbon vacancies in neutron-irradiated SiC: A high-field electron paramagnetic resonance study. *Phys. Rev. B* **67,** 125207 (2003).

12. Soltamov, V. A., Soltamova, A. A., Baranov, P. G. & Proskuryakov, I. I. Room Temperature Coherent Spin Alignment of Silicon Vacancies in 4H- and 6H-SiC. *Phys. Rev. Lett.* **108,** 226402 (2012).

13. Mizuochi, N. *et al.* Continuous-wave and pulsed EPR study of the negatively charged silicon vacancy with S=3/2 and $C_{3v}$ symmetry in n-type 4H-SiC. *Phys. Rev. B* **66,** 235202 (2002).

14. Cochrane, C. J., Lenahan, P. M. & Lelis, A. J. An electrically detected magnetic resonance study of performance limiting defects in SiC metal oxide semiconductor field effect transistors. *J. Appl. Phys.* **109,** 14506–14512 (2011).

15. Isoya, J. *et al.* EPR identification of intrinsic defects in SiC. *Phys. status solidi* **245,** 1298–1314 (2008).

16. Wimbauer, T., Meyer, B. K., Hofstaetter, A., Scharmann, A. & Overhof, H. Negatively charged Si vacancy in 4H SiC: A comparison between theory and experiment. *Phys. Rev. B* **56,** 7384–7388 (1997).

17. Yao, W., Liu, R.-B. & Sham, L. J. Theory of electron spin decoherence by interacting nuclear spins in a quantum dot. *Phys. Rev. B* **74,** 195301 (2006).

18. Zhao, N., Ho, S.-W. & Liu, R.-B. Decoherence and dynamical decoupling control of nitrogen vacancy center electron spins in nuclear spin baths. *Phys. Rev. B* **85,** 115303 (2012).

19. Yang, L.-P. *et al.* Electron Spin Decoherence in Silicon Carbide Nuclear Spin Bath. *Arxiv Prepr. arXiv1409.4646* (2014).

20. Chandrasekhar, S. Stochastic Problems in Physics and Astronomy. *Rev. Mod. Phys.* **15,** 1–89 (1943).

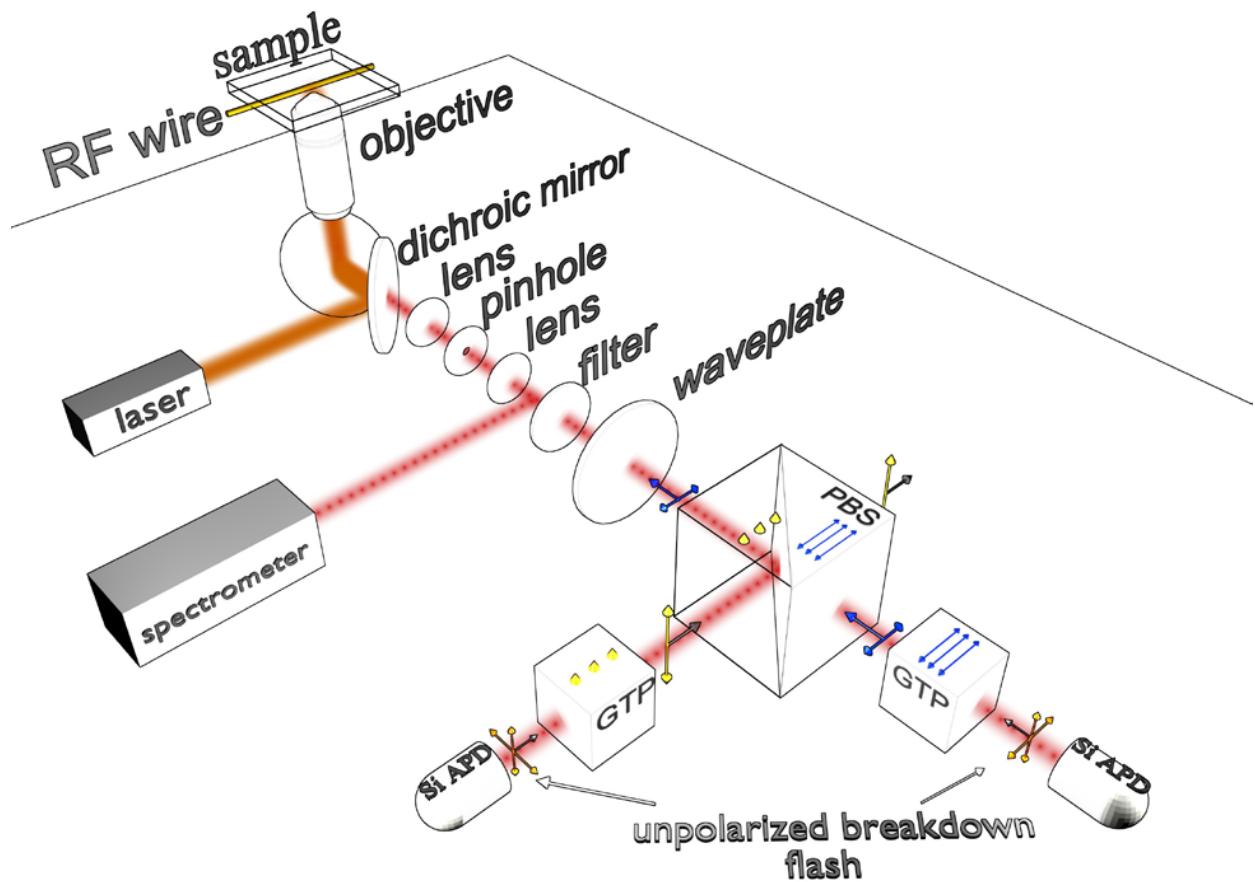

**Figure S1:** Experimental Setup. The excitation beam (orange) was focused with an oil objective on the sample. The fluorescent light from $T_{V2}$ centers (red) passes the dichroic mirror while laser light is blocked. To suppress background emission, the light from the sample was passed through a pinhole in the focus of two lenses (confocal configuration). This light was filtered by 800 nm long pass filter and sent to a spectrometer or 905 nm long pass filter and sent to the APDs. The usual 50:50 BS before the APDs was replaced with a PBS and a half-wave plate. The half wave plate was adjusted such that the fluorescent intensities detected by each APD become identical in order to establish HBT configuration. Blocking of the breakdown emissions from Si APDs was achieved by using both Glan-Tayler polarizers (GTP) before the APDs and the PBS. The breakdown flash from the APDs polarized either parallel or perpendicular to the optical table by the GTPs cannot enter the path to the other APDs due the PBS.

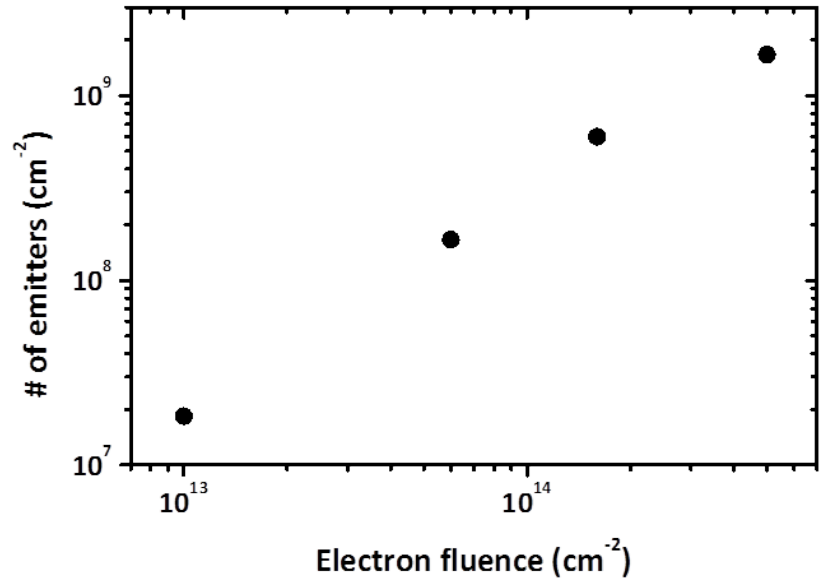

**Figure S2:** Concentration of $T_V$ centers as a function of the electron fluence.

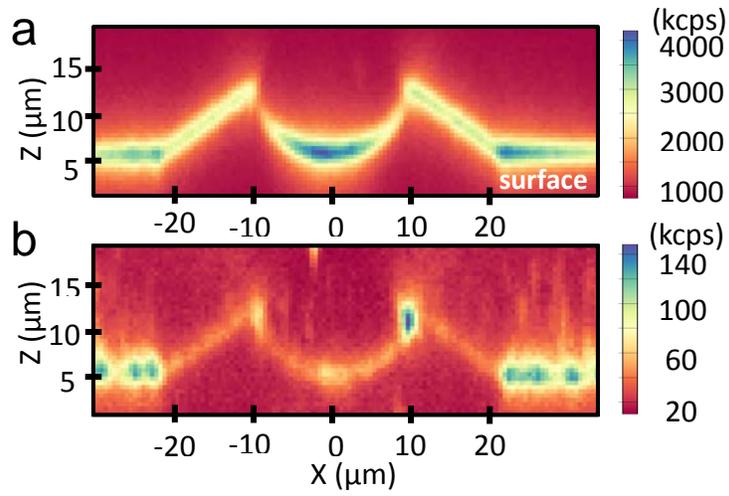

**Figure S3:** Confocal xz-scans with 12 mW at 660 nm excitation showing **a,** bright surface emissions up to 4000 kcps before etching, and **b,** drastically reduced surface emissions after etching.

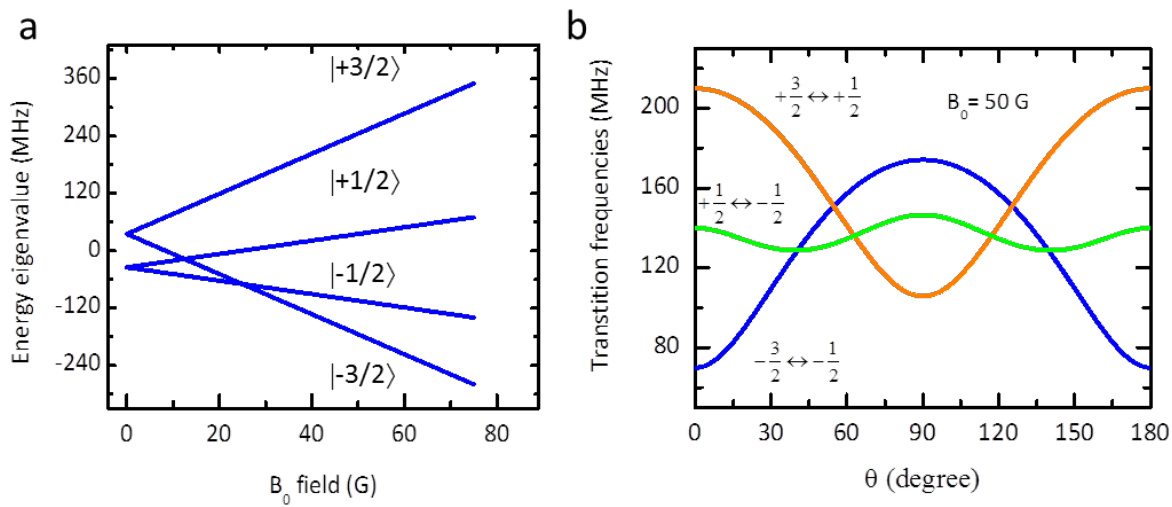

**Figure S4:** Calculated eigenvalues and transition frequencies of a S=3/2 system. **a,** Energy eigenvalues of each spin sublevel of a $T_{V2a}$ as a function of the axial $B_0$ field strength. **b,** Calculated resonance frequencies of the allowed transitions of a S=3/2 system as a function of the polar angle at $|B_0|$=50 G

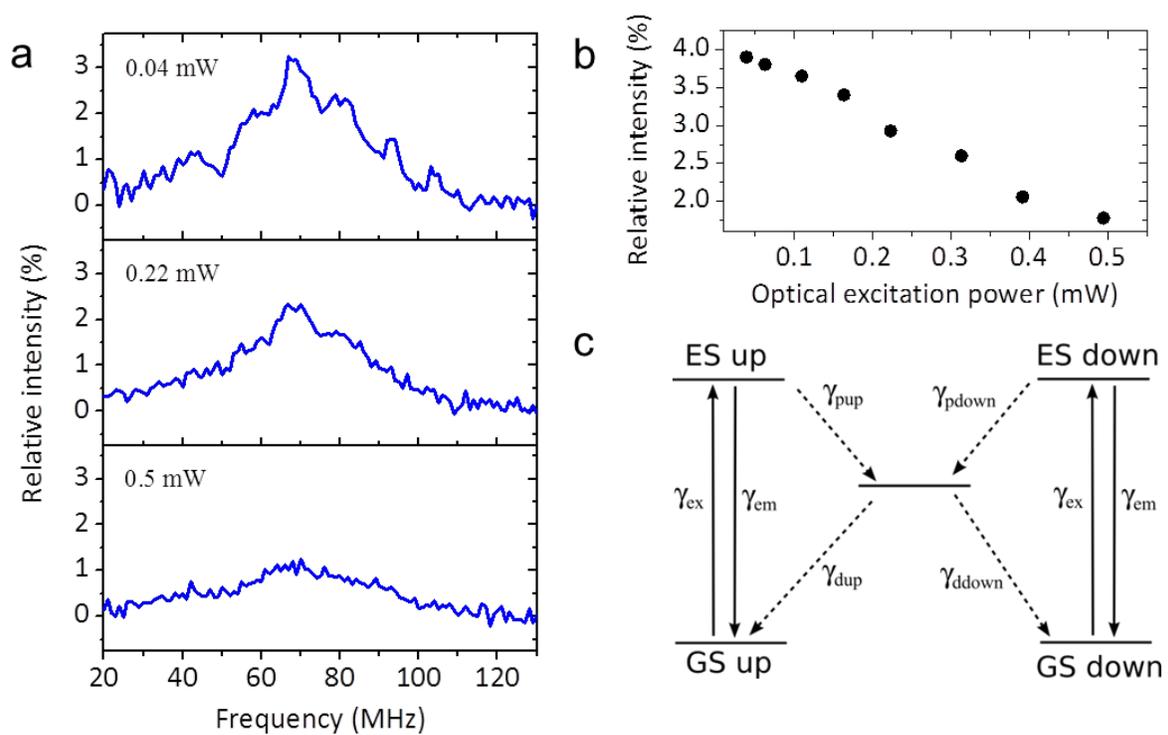

**Figure S5. a,** ODMR spectra of a single $T_{V2a}$ center measured at various optical excitation powers. **b,** Obtained spin signal intensity as a function of the optical excitation power. **c,** Rate model of the optical cycle.

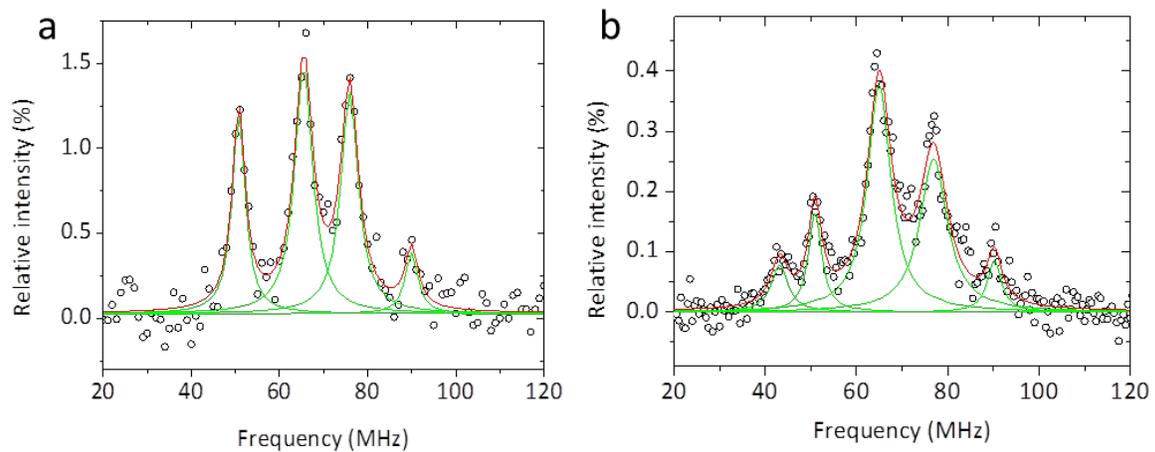

**Figure S6:** Fine structure of the zero-field ODMR spectra of $T_{V2a}$ centers: **a,** Single spin $T_{V2a}$ ODMR measured with weak $B_1$ field (FWHM ≈ 5 MHz) and 0.1 mW 730 nm laser excitation, filtered by 905 nm long pass filter. Sharp peaks are observable at 51, 65, 76, and 90 MHz. **b,** A $T_{V2a}$ ensemble ODMR spectrum at zero magnetic field with 10 mW 905 nm excitation and filtered by 925 nm long pass filter. Multiple peaks are found at the similar positions as in **a**, 51, 54, 77, and 90 MHz except for an additional small peak at 43 MHz. The ensemble sample is 4H-SiC bulk grown by high-temperature chemical vapour deposition having the concentration of the $T_{V2a}$ center in the mid $10^{15}$ cm$^{-3}$ range.

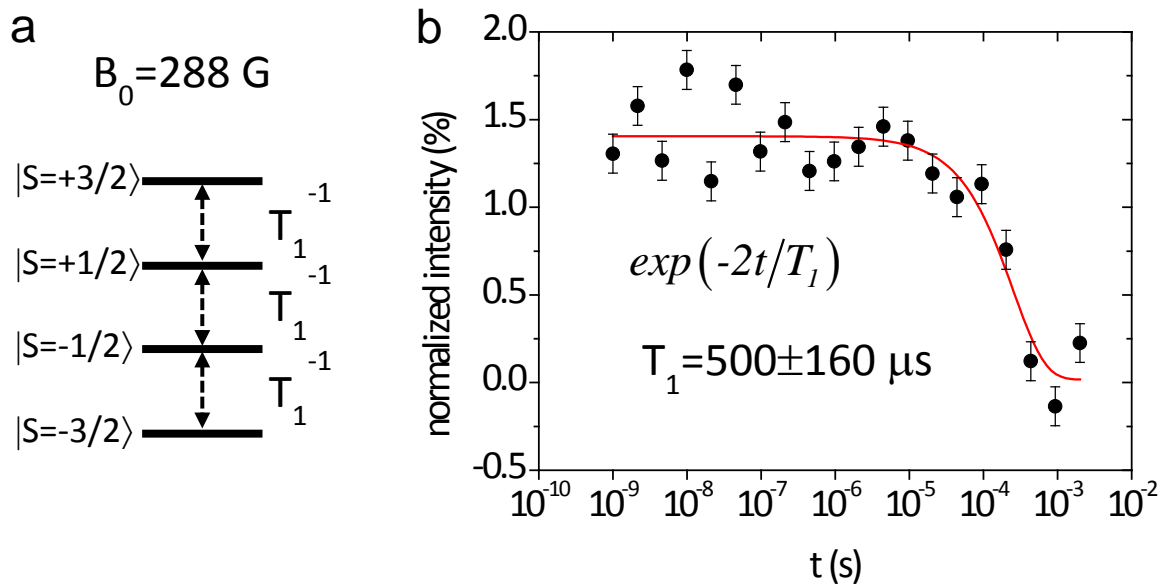

**Figure S7:** $T_1$ relaxations between the sublevels of the ground state of a $T_{V2a}$ center at axial $B_0$= 288 G: **a,** Rates describing the relaxations between each sublevels of the ground state after the steady states are established by a laser pulse. **b,** Obtained $T_1$ relaxation curve (see text). An exponential fitting (red curve) reveals $T_1 \approx 500$ µs.

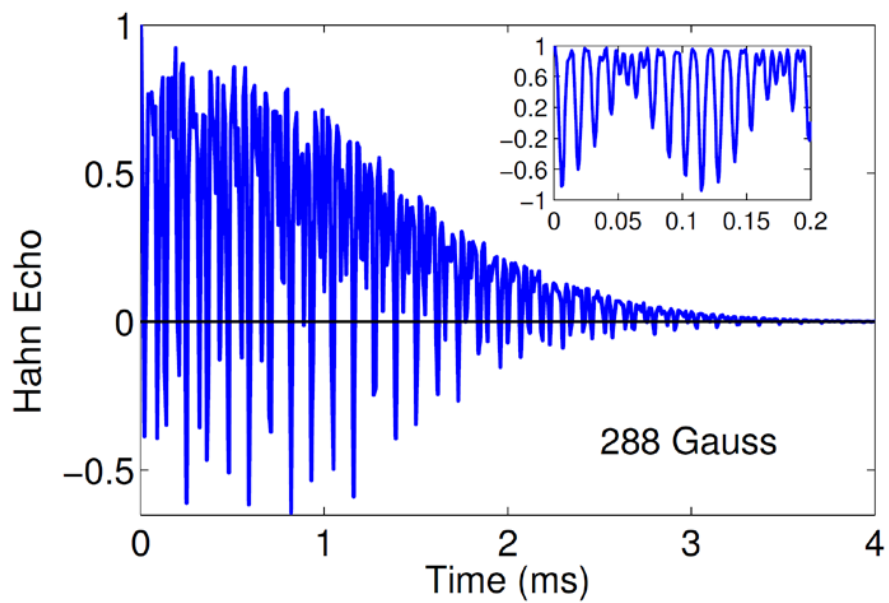

**Figure S8:** The numerical calculation of decoherence of a $T_{V2}$ center in 288 Gauss magnetic field. The inset shows the close-up of the modulations induced by the coherent coupling to a proximal $^{29}$Si nuclear spin.